**Monomer-induced customisation of UV-cured atelocollagen hydrogel networks**

He Liang,[1, 2] Stephen J. Russell,[1] David J. Wood,[2] Giuseppe Tronci[1,2*]

[1] Clothworkers' Centre for Textile Materials Innovation for Healthcare, School of Design, University of Leeds, United Kingdom

[2] Biomaterials and Tissue Engineering Research Group, School of Dentistry, St. James's University Hospital, University of Leeds, United Kingdom

* Email correspondence: G.Tronci@leeds.ac.uk

**Abstracts**

The covalent functionalisation of type I atelocollagen with either 4-vinylbenzyl or methacrylamide residues is presented as a simple synthetic strategy to achieve customisable, cell-friendly UV-cured hydrogel networks with widespread clinical applicability. Molecular parameters, i.e. the type of monomer, degree of atelocollagen functionalisation and UV-curing solution, have been systematically varied and their effect on gelation kinetics, swelling behaviour, elastic properties and enzymatic degradability investigated. UV-cured hydrogel networks deriving from atelocollagen precursors functionalised with equivalent molar content of 4-vinylbenzyl ($F_{4VBC}$= 18±1 mol.%) and methacrylamide ($F_{MA}$= 19±2 mol.%) adducts proved to display remarkably-different swelling ratio ($SR$= 1963±58–5202±401 wt.%), storage modulus ($G'$= 17±3–390±99 Pa) and collagenase resistance ($\mu_{rel}$= 18±5–56±5 wt.%), similarly to the case of UV-cured hydrogel networks obtained with the same type of methacrylamide adduct, but varied degree of functionalisation ($F_{MA}$= 19±2–88±1 mol.%). UV-induced network formation of 4VBC-functionalised atelocollagen molecules yielded hydrogels with increased stiffness and enzymatic stability, attributed to the molecular rigidity of resulting aromatised crosslinking segment, whilst no toxic response was observed with osteosarcoma G292 cells. Although to a lesser extent, the pH of the UV-curing solution also proved to affect macroscopic hydrogel properties, likely due to the altered organisation of atelocollagen molecules during network formation. By leveraging the knowledge gained with classic synthetic networks, this study highlights how the type of monomer can be conveniently exploited to realise customisable atelocollagen hydrogels for personalised medicine, whereby the structure-property relationships can be controlled to meet the requirements of unmet clinical applications.

## 1. Introduction

As a simple mimetic of the extracellular matrix (ECM) of biological tissues hydrogels have been widely applied in the biomedical field [1], for applications including regenerative medicine [2], wound care [3] and controlled drug delivery [4]. The high swelling in aqueous medium makes hydrogels inherently soft, so that molecular manipulation of their constituent building blocks [5] is key to enable customised -properties and functions for personalised medicine.

At the molecular scale, hydrogels typically consist of hydrophilic polymer networks, crosslinked via either physical or covalent linkages, whilst short amino acidic sequences have also been proven to generate water-swollen supramolecular structures [6]. Although various synthetic polymers have been employed [7,8,9,10,11,12,13], a great deal of attention has recently been paid to the use of natural, ECM-extracted, biopolymers as building blocks of multifunctional hydrogels [14,15]. The selection of biopolymer-based building blocks is attractive since it ensures that the resulting material has biomimetic features from the molecular up to the macroscopic scale, due to the presence of cell-binding sequences, fibre-forming biopolymer capability (with resulting fibre dimensions comparable to the ones found in the ECM) [16], and the water-swollen macroscopic state, respectively. In comparison to synthetic polymers, however, ensuring customisation and reliable structure-property relationships in biopolymer-based systems is highly challenging, due to the inherent batch-to-batch variability and the hardly-controllable secondary interactions between biopolymer chains [17,18]. Leveraging the knowledge gained with classic polymers, the synthesis of covalently-crosslinked hydrogel networks made of linear biopolymers, e.g. hyaluronic acid [19], has been successful aiming to achieve defined structure-property relationships [20]. Other than linear biopolymers, such a level of customisation is still hardly realised when employing biopolymers with increased organisational complexity such as collagen, because of the inherently-limited chemical accessibility and solubility. Customisation of these building blocks would on the other hand enable the formation of systems with superior biofunctionality and widespread clinical applicability. Type I collagen is one the most abundant structural proteins of connective tissues, and it constitutes the main organic component of bone, skin and tendon [21]. As the most abundant collagen in humans, type I collagen-based hydrogels have been widely employed for therapeutics and diagnostics, with multiple commercial products being routinely used in the clinic [22]. Other than its hierarchical, water-insoluble organisation found in biological tissues, collagen ex vivo is predominantly extracted in the form of a water-soluble triple helix, due to the extraction-induced breakdown of covalent crosslinks of collagen found in vivo. To enable applicability of the extracted water-soluble product in physiological conditions, restoration of in vivo-like covalent crosslinks ex vivo is a promising strategy. Reaction of side chain terminations with bifunctional segments, e.g. diisocyanates [23,24], diacids [25,26], and dialdehydes [27,28], or carbodiimide-induced intramolecular crosslinking [29,30] proved to induce some adjustment in the mechanical properties of resulting collagen materials. At the same time, the use of toxic reagents and the relatively-long reaction time prevent the application of these synthetic strategies for the delivery of cell-friendly, in-situ forming hydrogels [31] or to achieve material customisation at the bed side. Another challenge associated with above-mentioned strategies is that the enzymatic degradability of resulting collagen materials is still relatively quick, so that the requirements of specific clinical application, e.g. in guided bone regeneration, cannot be entirely fulfilled [32]. Ultimately, above-mentioned crosslinking strategies are often associated with unwanted side reactions, so that defined structure-property relationships are hardly-developed [20,23,27].

Rather than one-step crosslinking reactions, photoinduced network formations mechanisms have been recently investigated for the development of in situ-forming, cell-encapsulating collagen hydrogels [33]. Derivatisation of collagen triple helices with photoactive, e.g. methacrylamide, groups has been widely demonstrated to rapidly generate covalent networks via photoinduced free radical crosslinking [34] and UV-induced thiol-ene [35] click reactions. Although resulting systems proved to display varied storage modulus [34,36], collagen hydrogel customisation often relies on the incorporation of a synthetic copolymer, e.g. polyethylene glycol (PEG), in the crosslinked network. Whist this approach affords wide tailoring in macroscopic properties, incorporation of the synthetic phase may affect the biofunctionality of resulting system. In an effort to expand the customisability of collagen hydrogels and avoid the use of copolymers, functionalisation of collagen with 4-vinylbenzyl residues has recently been reported. UV-cured 4VBC-functionalised collagen triple helices proved to display significantly increased compression properties with respect to methacrylated variants [37], whereby the introduction of 4VBC aromatic rings was found to impact on the activity of matrix metalloproteinases (MMPs) in vitro [38,39]. Although insightful, these studies did not systematically investigate the effect of the type of monomer and respective degree of functionalisation on the macroscopic properties of resulting hydrogels, due to the limited chemical accessibility of the protein backbone. On the other hand, systematic investigations on the effect of covalently-coupled monomer and respective degree of functionalisation on network properties could open up new avenues aiming to develop simple synthetic routes yielding customisable collagen systems for personalised medicine.

This study therefore is focused on the synthesis of photoactive atelocollagen precursors and consequent UV-cured networks, whereby the effects of the (i) type of monomer, (ii) the degree of atelocollagen functionalisation, and (iii) the UV-curing aqueous solution were addressed. Pepsin-solubilised type I telopeptide-free collagen, i.e. atelocollagen, was selected as a purified, minimally-antigenic building block with comparable chemical composition and dichroic properties with respect to acid-extracted type I collagen [38,40]. Firstly, photoactive atelocollagen precursors with comparable molar content, but different type of photoactive adduct, i.e. either 4-vinylbenzyl or methacrylamide adduct, were considered. Secondly, photoactive atelocollagen precursors with varied molar content of the same type of photoactive, i.e. methacrylamide, adduct were addressed. Ultimately, the aqueous solution employed for the solubilisation of functionalised atelocollagen, i.e. hydrochloric acid, acetic acid and phosphate buffered saline solution, was also varied during network formation to investigate the effect of environmental conditions on resulting hydrogels. Hydrogel networks were prepared via UV-induced free radical crosslinking of photoactive atelocollagen precursors in the presence of 2-hydroxy-1-[4-(2-hydroxyethoxy) phenyl]-2-methyl-1-propanone (I2959), which was used as a water-soluble, cell-friendly photoinitiator [35]. Resulting UV-cured hydrogel networks were characterised with regards to their gelation kinetics, swelling and compression properties, enzymatic degradability, and cytotoxicity, aiming to establish defined structure-property relationships and systematic material customisation

## 2. Materials and methods

### 2.1 Materials

Pepsin-extracted type I bovine atelocollagen (AC, 6 mg·mL$^{-1}$) solutions in 10 mM hydrochloric acid (HCl) were purchased from Collagen Solutions PLC (Glasgow, UK). 4-vinylbenzyl chloride (4VBC), methacrylic anhydride (MA) and triethylamine (TEA) were purchased from Sigma-Aldrich. 2-Hydroxy-1-[4-(2-hydroxyethoxy) phenyl]-2-methylpropan-1-one (I2959) and deuterium oxide were purchased from Fluorochem Limited (Glossop, UK). Ninhydrin 99% was supplied by Alfa-Aesar (Massachusetts, USA). Polysorbate 20, absolute ethanol and diethyl ether were purchased from VWR internationals. All other chemicals were purchased from Sigma-Aldrich unless specified.

### 2.2 Synthesis of photoactive precursors

To achieve the MA-functionalised products, AC solutions (6 mg·mL$^{-1}$ in 10 mM HCl) were diluted to a concentration of 3 mg·mL$^{-1}$ via addition of 10 mM HCl and equilibrated to pH 7.5. MA and TEA were added at varied molar ratios with respect to the molar lysine content in AC ([MA][Lys]-1 = 0.1-25). When an MA/Lys molar ratio of 0.1-1 was selected, TEA was added with a 10 molar ratio with respect to the collagen lysine content. When an MA/Lys molar ratio of 25 was selected, an equimolar monomer content of TEA was added ([TEA]=[MA]). After 24 hours, the functionalisation reaction was stopped by precipitating the reacting mixture in 10-volume excess of absolute ethanol. Following at least 8-hour incubation in ethanol, the reacted, ethanol-precipitated product was recovered by centrifugation and air dried. The 4VBC-functionalised AC was prepared following the above protocol, whereby polysorbate 20 (PS-20) was added prior to the functionalisation reaction in order to mediate the solubility of 4VBC in water. Hence, the diluted (3 mg·mL$^{-1}$) and pH-equilibrated AC solution was supplemented with 1 wt.% PS-20 (with respect to the weight of the diluted AC solution) prior to addition of 4VBC and TEA at a fixed molar ratio of 25 ([4VBC]·[Lys]$^{-1}$= 25; [4VBC]=[TEA]).

### 2.3 Characterisation of reacted AC products

TNBS assay (n=3) was used to measure the derivatisation of amino to vinyl groups and respective degree of AC functionalisation, as previously reported [38-39]. Briefly, 11 mg of dry samples was mixed with 1 mL of 4 wt.% NaHCO$_3$ and 1 mL of 0.5 wt.% TNBS solution. The mixture was reacted at 40°C for 4 hours, following by addition 3 mL of 6 N HCl for one more hour to induce complete sample dissolution. The solution was then cooled down to room temperature, diluted with 5 mL of distilled water, and extracted (×3) with 15 mL diethyl ether to remove any non-reacted TNBS species. An aliquot of 5 mL was collected and diluted in 15 mL of distilled water and the reading was recorded using an UV-Vis spectrophotometer (Model 6305, Jenway) against the blank. The molar content of primary free amino groups (largely attributed to the side chains of lysine) was calculated via Equation 1:

$$\frac{mol(Lys)}{g(AC)} = \frac{2 \times Abs(346\ nm) \times 0.02}{1.46 \times 10^4 \times b \times x}$$ (Equation 1)

where *Abs (346 nm)* is the UV absorbance value recorded at 346 nm, *2* is the dilution factor, *0.02* is the volume of the sample solution (in litres), *1.46×10$^4$* is the molar absorption coefficient for 2,4,6-trinitrophenyl lysine (in M$^{-1}$ cm$^{-1}$), *b* is the cell path length (1 cm), and *x* is the weight of the dry sample. The degree of functionalisation (*F*) was calculated via Equation 2:

$$F = \left(1 - \frac{mol(Lys)_{funct.}}{mol(Lys)_{AC}}\right) \times 100$$ (Equation 2)

where *mol(Lys)$_{AC}$* and *mol(Lys)$_{funct.}$* represent the total molar content of free amino groups in native and functionalised atelocollagen, respectively.

Further confirmation of AC functionalisation was assessed via Ninhydrin assay (n=3). Briefly, 10 mg of the dry sample was mixed with 4 mL of distilled water and 1 mL of 8 wt.% Ninhydrin solution in acetone. The mixture was reacted at 100 °C for 15 mins and the reaction terminated by cooling in ice and adding 1 mL of 50% ethanol (v/v). The molar content of amino groups was measured by reading the absorbance at 570 nm against the blank. A standard curve calibration was prepared by measuring the known mass of collagen.

TNBS and ninhydrin assays were ultimately coupled with $^1$H-NMR spectroscopy (JEOL ECA, 600 MHz). $^1$H-NMR spectra of native, MA- and 4VBC-functionalised AC were recorded following dissolution of 5 mg of dry sample in 1 mL of 10 mM deuterium chloride.

## 2.4 Synthesis of UV-cured AC networks

Either MA- or 4VBC-functionalised AC products were dissolved at a fixed concentration of 0.8 wt.% in either 10 mM HCl (pH 2.1), 17.4 mM acetic acid (AcOH, pH 3.4) or 10 mM phosphate buffered saline (PBS, pH 7.5) solutions supplemented with 1 wt.% I2959 photo-initiator. Resulting AC solutions (0.8 wt.% functionalised AC, 0.992 wt.% I2959) were centrifuged at 3000 rpm for 5 mins to remove any air bubble and then cast onto a 24 well plate (Corning Costar, 0.8 g per well). Well plates were irradiated with a UV lamp (346 nm, 8 mW cm$^{-2}$, Spectroline) for 30 mins on both top and bottom side, leading to the formation of hydrogels. Irradiation intensities were measured with an International Light IL1400A radiometer equipped with a broadband silicon detector (model SEL033), a 10× attenuation neutral density filter (model QNDS1), and a quartz diffuser (model W). The UV-cured hydrogels were carefully removed from the plate and washed in distilled water (15 mins, ×3), followed by dehydration *via* an ascending series of ethanol and air drying.

## 2.5 Quantification of swelling ratio and gel content

Dry UV-cured samples (n=4) of known mass ($m_d$) were individually incubated in PBS (10 mM, pH 7.5, 1.5 mL) at room temperature for 24 hours. The swelling ratio (*SR)* was calculated by Equation 3:

$$SR = \frac{m_s - m_d}{m_s} \times 100 \qquad \textbf{(Equation 3)}$$

where $m_s$ is the mass of the PBS-equilibrated UV-cured sample.

The gel content (n=4) was measured to investigate the overall portion of the covalent hydrogel network insoluble in 17.4 mM acetic acid solution [39]. Dry collagen networks ($m_d$: 10 mg – 20 mg) were individually incubated in 2 mL of 17.4 mM AcOH for 24 hours. Resulting samples were further air dried and weighed. The gel content (*G*) was calculated by Equation 4:

$$G = \frac{m_1}{m_d} \times 100 \qquad \textbf{(Equation 4)}$$

where $m_1$ is the dry mass of collected samples.

## 2.6 Compression test

PBS-equilibrated hydrogel discs (Ø: 14 mm; h: 5-6 mm, n=3) were compressed at room temperature with a compression rate of 3 mm·min$^{-1}$ (Instron ElectroPuls E3000). A 250 N load cell was operated up to complete sample compression. Stress-compression curves were recorded and the compression modulus quantified as the slope of the plot linear region at 25-30% strain.

## 2.7 Degradation tests

Dry samples (n=4) of either UV-cured AC network or native AC were incubated for 4 days (37°C, pH 7.5) in 1 mL of 50 mM [tris(hydroxymethyl)-methyl-2-aminoethane sulfonate] (TES) buffer containing 0.36 mM calcium chloride and supplemented with 5 CDU of collagenase type I from *Clostridium histolyticum* (125 CDU·mg$^{-1}$). Following 4-day incubation, the samples were washed in distilled water, dehydrated via an ascending series of ethanol solutions and air dried. The relative mass ($\mu_{rel}$) of samples was determined according to Equation 5:

$$\mu_{rel} = \frac{m_4}{m_d} \times 100 \qquad \textbf{(Equation 5)}$$

where $m_4$ and $m_d$ are the masses of the dry partially-degraded and dry freshly-synthesised samples, respectively.

**2.8 UV-curing rheological measurements**

The UV-induced kinetics of network formation (n=3) was measured by a modular compact rotational rheometer (MCR 302, Anton Paar, Austria) equipped with a UV curing module (Ominicure 1500, Excelitus Technologies). Functionalised atelocollagen products (0.8 wt.%) were dissolved in I2959-supplemented aqueous solutions (1 wt.% I2959), followed by time sweep measurement at a strain of 0.1 % and frequency of 1 Hz. Values of storage ($G'$) and loss ($G''$) modulus were recorded during time sweep measurements under irradiation with UV light. The oscillatory shear was applied to a transparent glass parallel plate (Ø 25 mm) and the gap between the plates was 300 μm. UV light (365 nm, 8 mW·cm$^{-2}$) was initiated after 5 s of shear oscillation at 21°C. The gelation time ($\tau$) was determined by the temporal interval between the UV activation and complete gelation.

**2.9 Cell culture study**

G292 cells were cultured in Dulbecco's modified Eagle's medium (DMEM), supplemented with 10% fetal bovine serum (FBS), 1% glutamine, and 2.5 mg·mL$^{-1}$ penicillin–streptomycin, in a humidified incubator at 37 ºC and 5% $CO_2$. Cells were passaged every 3 days with 0.25% trypsin/0.02% EDTA. UV-cured samples were individually synthesised on to a 24-well plate, extensively washed in distilled water and dehydrated in an ascending series of ethanol-distilled water (0, 20, 40, 60, 80, (3×) 100 vol.% EtOH) to remove any acidic or ethanol residues. Prior to cell seeding, hydrogels were disinfected in a 70 vol.% ethanol solution under UV light and washed in PBS (3×, 10 min) to remove any acidic or ethanol residues. G292 cells (8·10$^3$ cells·mL$^{-1}$) were seeded on top of the sample (following UV disinfection) and incubated at 37 ºC for up to 7 days. After incubation, samples (n = 6) were washed with PBS (×3) and transferred to a new 24-well plate before adding the dying agent of Calcein AM and Ethidium homodimer-1. The sample plate was then incubated for 20 minutes away from light. Live /dead stained hydrogels were placed onto a glass slide for fluorescence microscopy imaging (Leica DMI6000 B).Cells grown on tissue culture treated plastics were used as positive control (Nunc, UK). Other than live/dead staining, cell viability was assessed at selected time points using Alamar Blue assay (ThermoFisher Scientific, UK) according to the manufacturer's guidance. VP-SEM (Hitachi S-3400N VP) combined with Deben cool stage control (Model: LT3299) was also employed for high resolution imaging of cell attachment on hydrated samples after 7-day incubation under low pressure (60-70 Pa).

**2.10 Statistical analysis**

Data are presented as mean ± standard deviation (SD). Statistical analysis was carried out with the Student's *t*-test. A *p* value lower than 0.05 was considered to be significantly different.

**3. Results and discussion**

In the following, the synthesis and characterisation of UV-cured hydrogel networks made of either 4-vinylbenzylated or methacrylated atelocollagen is presented. The effects of covalently-coupled monomer, degree of atelocollagen functionalisation and UV-curing aqueous solution was investigated, aiming to draw controlled structure-property relationships. The synthesis of the covalent networks was linked to the degree of functionalisation of respective photoactive AC solutions, whilst respective UV-induced gelation kinetics was assessed in either acidic or basic conditions. Resulting UV-cured networks were characterised with regards to rheological properties, gel content, swelling ratio, compression modulus, enzymatic degradability and cellular tolerability.

The sample nomenclature is as follows: functionalised AC samples are coded as 'XXXYY', where 'XXX' identifies the type of monomer introduced on to the AC backbone, i.e. either 4VBC or MA; and 'YY' describes the monomer/Lys molar ratio used in the functionalisation reaction. UV-cured samples are coded as 'XXXYY(Z)*', where 'XXX' and 'YY' have the same meaning as previously discussed; 'Z' indicates the UV-curing aqueous solution used to dissolve the functionalised AC sample, i.e. either 10 mM HCl (H), 17.4 mM acetic acid (A) or 10 mM PBS (P); '*' identifies a UV-cured hydrogel sample.

### 3.1 Synthesis of functionalised AC precursors

Following reaction with either 4VBC or MA, the degree of functionalisation ($F$) consequent to the covalent coupling of photoactive adducts on to the AC backbone was determined. Since network formation was pursed via UV-induced free radical crosslinking mechanism, the introduction of photoactive adducts was expected to be directly related to the crosslink density and macroscopic properties of respective AC networks. The reaction of AC with selected monomers proceeds via lysine-initiated nucleophilic substitution (Figure S1, Supp. Inf.), leading to the consumption of free amino groups and derivatisation with either 4-vinylbenzyl or methacrylamide adducts. TNBS [41] and ninhydrin [42] assays are two colorimetric assays widely employed for the determination of amino groups in proteins. Both assays have recently been employed for the characterisation of reacted gelatin [43,44] and acid-solubilised collagen [36,37,38,39] products, proving to correlate well with $^1$H-NMR spectroscopy. Both TNBS and ninhydrin assays were therefore selected in this study to measure the molar content of free amino groups in both native and reacted pepsin-solubilised AC samples, so that $F$ could be indirectly quantified.

An overall content of primary amino groups of $2.89 \cdot 10^{-4}$ mol·g$^{-1}$ was recorded on native atelocollagen via TNBS assay. Both colorimetric methods revealed a comparable and decreased molar content of amino groups in both MA- and 4VBC-reacted atelocollagen samples (Table 1), in line with previous reports [38,45]; whilst monomer-related geminal protons could not be clearly detected in $^1$H-NMR spectra of sample MA0.3 and 4VBC25 due to the overlapping with AC species (Figure S2, Supp. Inf.). The range of MA/Lys molar ratio selected during the functionalisation reaction proved to directly impact on the degree of atelocollagen methacrylation ($F_{MA}$= 4±1–88±1 mol.%), whilst an insignificant variation in the molar content of covalently-coupled 4VBC adducts ($F_{4VBC}$= 18±1 mol.%) was observed in respective products (data not shown) The different trend in degree of functionalisation observed in MA- with respect to 4VBC-reacted samples is attributed to the decreased reactivity and solubility of 4VBC in aqueous environment, as reported in previous publications [37], which was partially mitigated in this study by the addition of PS-20 in the reacting mixture.

Despite the different reactivity of MA with respect to 4VBC, the range of $F$ observed in resulting MA-functionalised AC is in line with the results reported in previous publications [33,34,36] with collagen-based materials.

**Table 1.** TNBS assay quantification of the amino group molar content and degree of functionalisation ($F$) in atelocollagen products (n=2) following reaction with either 4VBC or MA at varied monomer/Lys molar ratios. The quantification of amino group molar content was confirmed via Ninydrin assay (n=2). n,a.: not applicable.

| Sample ID | Molar ratio / [Monomer][Lys]$^{-1}$ | Amine groups /mol·g$^{-1}$ (×10$^{-4}$) | | $F$ /mol.% [1] |
| --- | --- | --- | --- | --- |
| | | TNBS | Ninhydrin | |
| **4VBC25** | 25 | 2.38 ± 0.02 | 2.38 ± 0.01 | 18 ± 1 |
| **MA0.1** | 0.1 | 2.77 ± 0.03 | n.a. | 4 ± 1 |
| **MA0.3** | 0.3 | 2.34 ± 0.06 | 2.40 ± 0.13 | 19 ± 2 |
| **MA0.5** | 0.5 | 2.07 ± 0.14 | n.a. | 28 ± 5 |
| **MA1** | 1 | 1.63 ± 0.11 | n.a. | 44 ± 4 |
| **MA25** | 25 | 0.36 ± 0.01 | 0.44 ± 0.04 | 88 ± 1 |

[1] $F$ was calculated considering an overall molar content of amino groups of 2.89×10$^{-4}$ mol·g$^{-1}$ in native AC, as revealed by TNBS assay.

Billiet et al. analysed the covalent coupling of (meth)acrylic anhydride to bovine type B gelatin and achieved up to ca. 80 mol.% methacrylation [44]. Ravichandran et al. reported a degree of porcine collagen methacrylation of 57-87 mol.%, and similar results ($F$= 39-86 mol.%) were obtained by Chaikof et al. with acid-soluble rat tail collagen [33]. Compared to the latter case, the results presented in this work with pepsin-soluble collagen suggest that the removal of telopeptides does not significantly impact on the molar content of free amino groups and functionalisation capability of resulting atelocollagen product, in line with a previous report [38].

In order to systematically investigate how above-mentioned differences at the molecular scale of AC precursors were reflected on the macroscale of resulting UV-cured networks, it was therefore of interest to systematically vary either the type of covalently-coupled monomer (whilst keeping respective degree of AC functionalisation constant) or the degree of functionalisation (with the same type of covalently-coupled monomer). Accordingly, samples

MA0.3 ($F_{MA}$ = 19 ± 2 mol.%), 4VBC25 ($F_{4VBC}$ = 18 ± 1 mol.%), and MA25 ($F_{MA}$ = 88 ± 1 mol.%) were selected for further investigations, and dissolved in either hydrochloric acid, acetic acid or PBS, as commonly-used solvent for collagen, prior to UV curing.

### 3.2 UV-induced network formation and time-sweep photorheometry

AC samples functionalised with methacrylamide residues (Figure 1 A) formed a clear solution in both acidic and PBS solutions, whilst sample 4VBC25 (Figure 1 B) could only be dissolved in 10 mM HCl (pH 2.1) and 17.4 mM AcOH (pH 3.4). The insolubility of the 4VBC-functionalised AC product in PBS agrees with previous reports [46] and may be attributed to secondary interactions developed between aromatic 4-vinylbenzyl residues following atelocollagen functionalisation and sample drying. Aromatic rings can mediate $\pi$-$\pi$ stacking interactions as well as act as hydrogen bond acceptor in aqueous environment [47]. Resulting $\pi$-$\pi$ stacking interactions developed between 4VBC-functionalised AC molecules are likely to be broken down at decreased rather than basic pH, supporting the observed complete dissolution of sample 4VBC25 in both HCl and AcOH solutions.

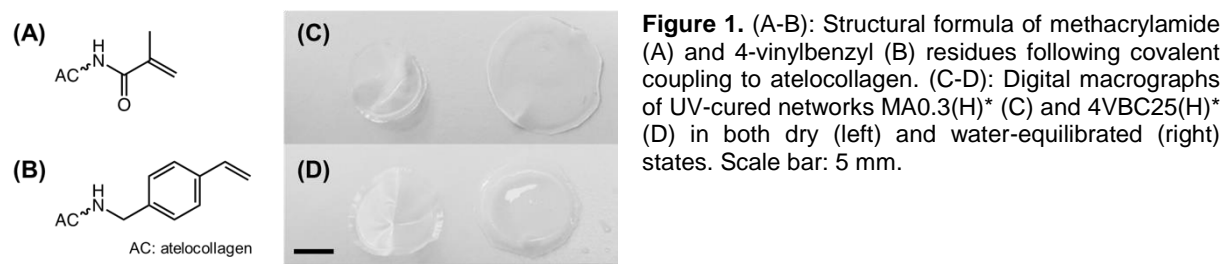

**Figure 1.** (A-B): Structural formula of methacrylamide (A) and 4-vinylbenzyl (B) residues following covalent coupling to atelocollagen. (C-D): Digital macrographs of UV-cured networks MA0.3(H)* (C) and 4VBC25(H)* (D) in both dry (left) and water-equilibrated (right) states. Scale bar: 5 mm.

All photoactive solutions successfully yielded hydrogel networks when treated with UV light, irrespective of the type of monomer introduced on to AC (Figure 1 C-D). To confirm the formation of a UV-cured hydrogel network and to explore the effect of methacrylamide and 4-vinlybenzyl residues on gelation kinetics, time sweep photorheometry was carried out prior to and during UV irradiation of AC photoactive solutions, and changes in viscosity as well as storage (*G'*) and loss (*G''*) modulus recorded. Prompt increase of *G'* and *G''* was observed following activation of the atelocollagen mixtures with UV light, in contrast to the case of the same sample being tested in the absence of UV light (Figure 2). Comparable values of *G'* and *G''* were also measured in the latter case, providing indirect evidence of the absence of a crosslinked network at the molecular level (Table 2). Both the gelation, i.e. the time required for the storage and loss moduli to equate, and plateau in storage modulus were reached within 12 s and 180 s, respectively (Table 2). The photoinitiator concentration (1 wt.%) selected to prepare gel-forming AC solutions proved key to reduce gelation time and to generate hydrogels with increased elastic modulus with respect to gel-forming AC solutions supplemented with decreased photoinitiator concentration (0.5 wt.%) (Figure 2B and C). These results, together with the fact that the final values of *G'* were significantly higher compared to the values of *G"* (Table 2), successfully indicate complete conversion of the photoactive solution into a UV-cured covalently-crosslinked network [48]. The obtained gelation kinetics curves are comparable to the ones described by thiol-ene collagen-poly(ethylene glycol) mixtures [35], whereby gel points were recorded within ~7 s of UV activation. This observation is interesting given that the orthogonality and oxygen-insensitivity of the UV-induced thiol-ene, with respect to the free-radical, crosslinking reaction, should be corelated with decreased gelation times [49].

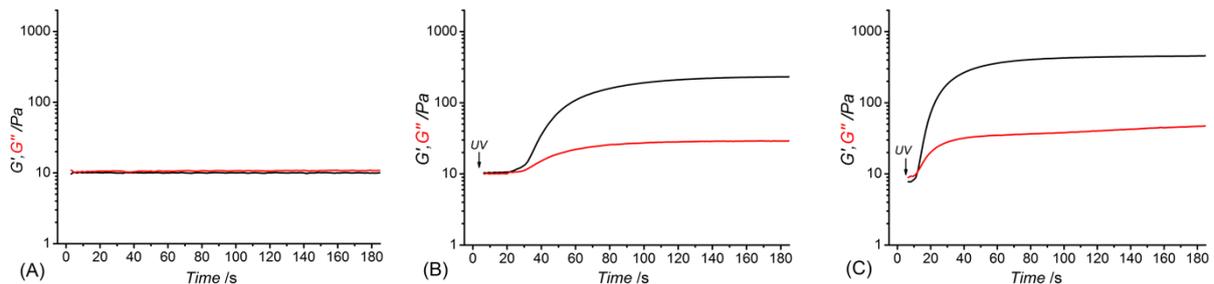

**Figure 2.** Rheograms of functionalised atelocollagen solution (0.8 wt.%) of sample 4VBC25 in 17.4 mM acetic acid supplemented with either 1 wt.% I2959 (A, C) or 0.5 wt.% I2959 (B). (A): Storage (*G'*, black) and loss (*G"*, red) modulus recorded in the absence of UV light. (B-C): UV-induced gelation kinetics of functionalised atelocollagen solutions. UV light was activated 5 s following shear oscillation.

Figure 3 and Table 2 describe the gelation kinetic profiles of samples 4VBC25, MA0.3 and MA25 when dissolved in either 10 mM HCl, 17.4 mM AcOH or 10 mM PBS solutions supplemented with 1 wt.% I2959, whilst an atelocollagen control sample was also tested following solubilisation in a 17.4 mM acetic acid solution supplemented with 1 wt.% I2959. Overall, sample MA25 was found to generate networks with the highest loss and storage modulus (*G'*= 618±87–18897±4793 Pa) and fastest gelation kinetics ($\tau$ ~ 6 s), followed by samples 4VBC25 ($G'$= 73±22–390±99 Pa) and MA0.3 ($G'$= 17±3–85±1 Pa). In contrast, only a marginal increase of $G'$ was observed following UV activation of the control atelocollagen solution, likely due to the much lower radical-induced crosslinking of amino acidic residues [50] with respect to the complete UV-induced network formation achieved with functionalised atelocollagen precursors. This result therefore demonstrates the importance of selected functionalisation routes in ensuring the synthesis of full covalent networks and provides additional evidence of the covalent coupling of selected monomers to atelocollagen. Other than the variation in elastic modulus, the viscosity of respective network-forming AC solutions was also recorded, whereby the sample MA25 proved to generate highly viscous solutions in all solutions investigated ($\eta$= 920±100–1400±190 mPa·s).

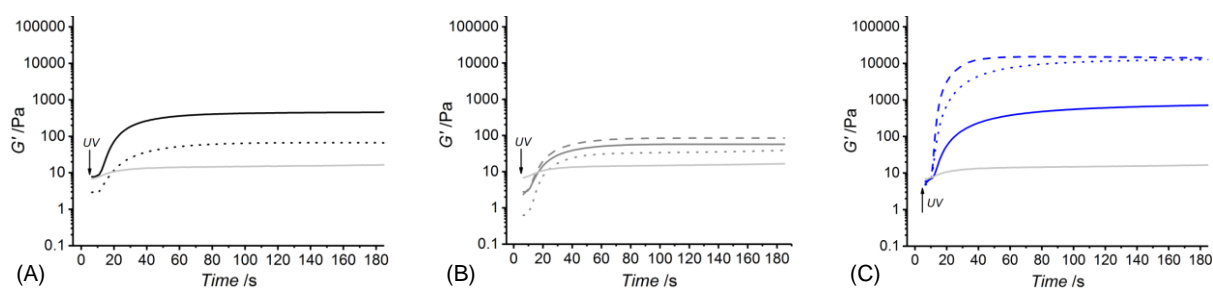

**Figure 3.** Kinetics of network formation studied via UV-equipped rheometry on samples 4VBC25 (A), MA0.3 (B) and MA25 (C), when dissolved (0.8 wt.%) in either 10 mM hydrochloric acid, 17.4 mM acetic acid or 10 mM PBS solution (each supplemented with 1 wt.% I2959). UV light was activated following 5 s shear oscillation. (A): (···): 4VBC25(H)*; (—): 4VBC25(A)*. (B): (···): MA0.3(H)*; (—): MA0.3(A)*; (– – –): MA0.3(P)*. (C): (···): MA25(H)*; (—): MA25(A)*; (– – –): MA25(P)*. A solution of atelocollagen (0.8 wt.%) in 17.4 mM acetic acid (supplemented with 1 wt.% I2959) was also tested and respective $G'$ profile reported in each plot (light grey curve).

The variations observed in loss and storage modulus and gelation kinetics support the trends in degree of functionalisation recorded at the molecular scale of the atelocollagen precursors, whereby sample MA25 exhibited the highest molar content of photoactive residues with respect to the other two samples. Given that the covalently-coupled photoactive adducts mediate the UV-induced free-radical crosslinking reaction, the increased storage and loss moduli as well as decreased gelation time measured in these samples therefore provide indirect confirmation that changes in the degree of AC functionalisation directly impact on the crosslink density, gelation kinetics and elastic properties of resulting UV-cured networks. The increased solution viscosity recorded with MA25 also correlates with respective degree of methacrylation, due to the increased molar content and molecular weight of derivatised lysine terminations.

Together with the effect of varied degrees of functionalisation, equimolar functionalisation of AC with distinct photoactive adducts, i.e. either methacrylamide or 4-vinylbenzyl residues, was also found to play a detectable effect on the macroscopic properties of resulting UV-cured hydrogels (Table 2). Both networks 4VBC25(H)* and 4VBC25(A)* displayed significantly increased storage and loss moduli with respect to the methacrylated equivalents (prepared in the same solvents). Also in this case, the trend in mechanical properties was supported by direct variations in respective network-forming solutions, whereby significantly-increased solution viscosity and decreased gelation times were measured with sample 4VBC25 ($\eta=$ 900±70–1820±80 mPa·s; $\tau=$ 8±0–10±0 s) compared to sample MA0.3 ($\eta=$ 240±20–390±20 mPa·s; $\tau=$ 11±1–12±1 s).

**Table 2.** Viscosity ($\eta$) and gelation time ($\tau$) of hydrogel-forming solution as well as storage ($G'$) and loss ($G''$) moduli of UV-cured hydrogel networks following complete gelation. Data are presented as mean ± SD (n=3).

| Sample ID | $\eta$ /mPa·s [1] | $G'_{max}$ /Pa | $G''_{max}$ /Pa | $\tau$ /s |
|---|---|---|---|---|
| AC[2] | 1777 ± 20 | 16 ± 1 | 13 ± 1 | 22 ± 5 |
| 4VBC25(H)* | 900 ± 70 [a] | 73 ± 22 [b] | 20 ± 3 [b] | 10 ± 0 [a] [d] [c] |
| 4VBC25(A)* | 1820 ± 80 | 390 ± 99 | 42 ± 7 | 8 ± 0 [a] |
| MA0.3(H)* | 390 ± 20 [a] [c] | 17 ± 3 [a] | 8 ± 5 [a] | 12 ± 1 [d] [c] |
| MA0.3(A)* | 240 ± 20 [a] | 58 ± 5 [a] | 17 ± 3 [d] | 11 ± 1 |
| MA0.3(P)* | 240 ± 20 [c] | 85 ± 1 [a] | 24 ± 2 [a] [d] | 11 ± 1 |
| MA25(H)* | 1200 ± 130 | 12967 ± 265 [a] | 42 ± 4 [a] [b] | 6 ± 1 [c] |
| MA25(A)* | 920 ± 100 [d] | 618 ± 87 [a] [b] | 8 ± 1 [a] [c] | n.a. |
| MA25(P)* | 1400 ± 190 | 18897 ± 4793 [b] | 146 ± 36 [b] [c] | 6 ± 0 |

[1] Viscosity of hydrogel-forming AC solutions prior to UV light activation
[2] Atelocollagen (0.8 wt.%) solution control prepared in 17.4 mM acetic acid supplemented with 1 wt.% I2959
[a] & [c]: $p <0.001$; [b]: $p <0.01$; [d]: $p <0.05$

The significant difference in storage and loss modulus measured in UV-cured aromatised atelocollagen networks with respect to equivalent methacrylated variants gives evidence of the major role played by the monomer covalently-coupled to, and resulting photocrosslinked segment between, AC molecules. Aromatic interactions have been exploited for the formation of physically-crosslinked self-assembled peptides [51], and for the controlled delivery of hydrophobic drugs [52]. The above-mentioned increased viscosity, storage modulus and loss modulus measured on 4VBC-functionalised atelocollagen solutions and corresponding UV-cured networks can mostly be attributed to $\pi$-$\pi$ stacking interaction capability and increased molecular rigidity of aromatic 4VBC moieties and consequent UV-cured aromatised segment, respectively. Other than that, the decreased gelation time recorded in solutions of sample 4VBC25 with respect to solutions of sample MA0.3 seems to correlate with the increased segment length of the 4-vinylbenzyl, with respect to methacrylamide, moiety (Figure 1 A), so that crosslinking-hindering steric effects are less likely in the former compared to the latter system. Rather than the typical variation of the molar content of photoactive groups, these results therefore demonstrate the possibility of adjusting the mechanical properties of atelocollagen hydrogels by simply varying the type of monomer covalently-coupled to the AC backbone.

Other than the degree and type of functionalisation, the effect of the UV-curing aqueous solution was also explored as an additional parameter to control material behaviour (Figure 3, Table 2). Hydrochloric acid solutions (10 mM, pH= 2.1) of AC precursors functionalised with equimolar content of either 4VBC or MA adducts yielded crosslinked hydrogels with the lowest storage ($G'_{4VBC25}$= 73±22 Pa; $G'_{MA0.3}$= 17±3 Pa) and loss ($G''_{4VBC25}$= 20±3 Pa; $G''_{MA0.3}$= 8±5 Pa) moduli, compared to hydrogels prepared from atelocollagen solutions in acetic acid (17.4 mM, pH 3.4) and PBS (10 mM, pH 7.5). Interestingly, similar trends were observed in the case of 4VBC-functionalised AC solutions, whereby significantly-increased viscosities were measured when the photoactive precursor was dissolved in hydrochloric acid ($\eta$= 900±70 mPa·s) compared to acetic acid ($\eta$= 1820±80 mPa·s) solutions (Table 2), whilst an opposite trend was observed in acidic solutions prepared with samples MA0.3 and MA25.

The solution-induced effect observed in AC products functionalised with equivalent molar content of methacrylamide and 4-vinylbenzyl chloride is in agreement with the results reported by Ratanavaraporn et al. [53], whereby increased solution viscosity and hydrogel compressive modulus were measured when native type I collagen was dissolved in aqueous solutions of decreased acidity. It is well known that the molecular organisation of native collagen molecules can be altered depending on environmental factors, such as pH, ionic strength and salt concentration, given their effect on the ionisation of amino acidic terminations and consequent secondary interactions [17,21,22,30]. The above-mentioned variations in network elastic properties suggest that a similar effect can still be observed in the case of functionalised atelocollagen with decreased $F$. Whilst fibrillogenesis was not expected during network formation ($T$< 37 ºC), the electrostatic repulsion of native atelocollagen molecules is expected to be increased in solutions of increased acidity [53,54]. Obviously, the degree and type of monomer adds further complexity to the solution-induced variability of protein organisation, given that the atelocollagen functionalisation is achieved by the consumption of ionisable lysine terminations and that selected covalently-coupled adducts can mediate further secondary interactions, i.e. $\pi$-$\pi$ stacking interactions and hydrogen bonds. The opposite pH-induced variation in solution viscosity measured with samples 4VBC25, on the one hand, and MA0.3 and MA25, on the other hand, reflects these considerations, given that aromatic interactions between 4-vinylbenzyl moieties are likely to be decreased at increased pH, whilst an opposite trend is expected with regards to the hydrogen bonding capability of methacrylamide residues. Whilst a solution pH-induced effect was clearly observed in hydrochloric and acetic acid solutions, the presence of salts in the PBS-based UV-curing system was likely to alter the capability of covalently-coupled moieties to mediate secondary interactions, explaining why no direct relationships between solution pH, on the one hand, and solution viscosity and hydrogel properties, on the other hand, could be observed.

**3.3 Swelling, gel content and compression properties**

Following the characterisation of functionalised AC precursors and gelation kinetics, the gel content ($G$), swelling ratio ($SR$) and compressive properties were quantified to further elucidate the molecular architecture and assess the structure-property relationships of obtained UV-cured atelocollagen networks.

All UV-cured networks displayed an averaged gel content of at least 80 wt.%, with the exception of sample MA0.3(H)* ($G$= 56±13 wt.%) (Figure 4). Samples MA25* displayed the smallest variation in gel content when prepared in either HCl, AcOH or PBS solutions, whilst the effect of the UV-curing solution was more visible in samples of MA0.3* and 4VBC25*. The high, and small solution-induced variation of, gel content measured in samples MA25* confirms that precursors with increased degree of functionalisation generate highly crosslinked network, regardless of the type of UV-curing solution and solution-induced secondary interactions, as indicated previously (Table 1 and 2).

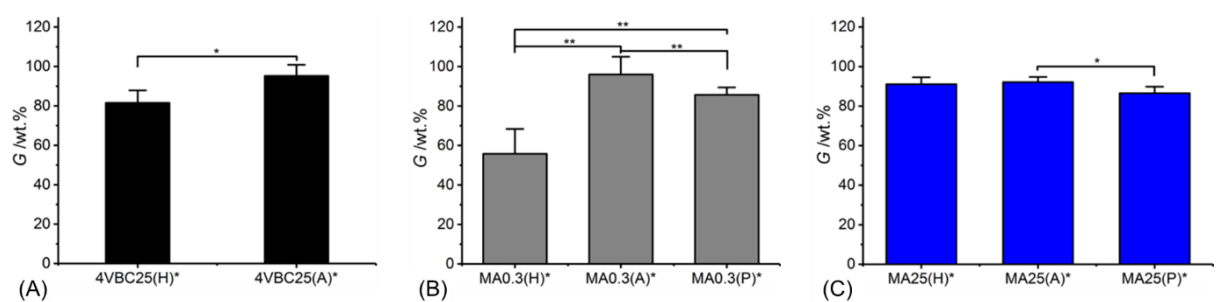

**Figure 4.** Gel content ($G$) of collagen networks prepared with varied monomers, monomer/Lys molar ratios and aqueous solutions. Data are shown as mean ± SD (n=4). ***: $p<0.001$; **: $p<0.01$; *: $p<0.05$.

In comparison to sample MA25*, networks 4VBC25(H)* and MA0.3(H)* prepared in HCl solutions revealed the lowest gel content within the same sample group, whilst increased $G$ values were measured in networks prepared in either AcOH, i.e. 4VBC25(A)* and MA0.3(A)*, or PBS, i.e. MA0.3(P)*. Given the comparable degree of functionalisation in the network precursors (Table 1), these solution-induced variations in gel content seem to correlate with the above-mentioned results in UV-curing solution viscosity (Table 2), whereby an increase of solution pH from 2.1 (in 10 mM HCl) to 3.4 (in 17.4 mM AcOH) proved to generate hydrogels 4VBC25* with increased storage modulus (Table 2). An increase in solution viscosity is likely to be associated with an increased vicinity between atelocollagen molecules and respective crosslink-forming photoactive residues, so that increased chances of free-radical crosslinking reaction and minimised steric hindrance effects can be expected. Similar effects were also reported with collagen hydrogels prepared from solutions with decreased viscosity, whereby decreased mechanical properties and increased pore size was observed [53].

Previously-discussed results in gel content proved to support the trends in swelling ratio (Figure 5), whereby UV-cured networks described an inverse relationship between *G* and *SR*. For example, samples 4VBC25(H)* (*SR*= 3769±111 wt.%) and MA0.3(H)* (*SR*= 5202±401 wt.%) displayed the highest swelling ratio, confirming the higher water uptake capability of these materials with respect to sample variants prepared in AcOH solutions, reflecting above-discussed gel content trends.

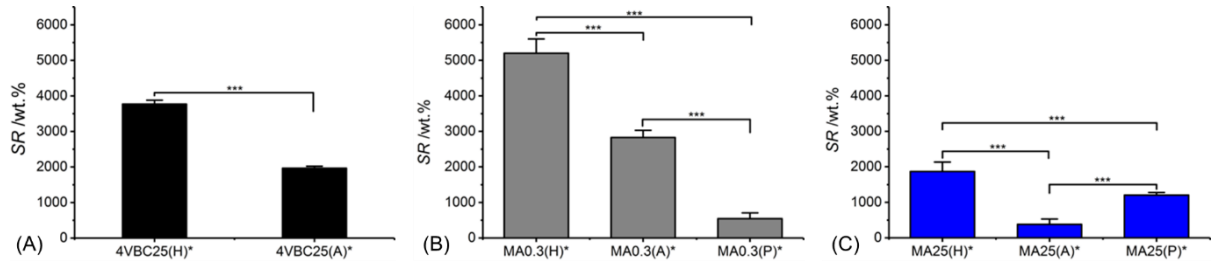

**Figure 5.** Swelling ratio (*SR*) in collagen networks prepared with varied monomers, monomer/Lys molar ratios and aqueous solutions. Data are shown as mean ± SD (n=4). ***: $p<0.001$; **: $p<0.01$; *: $p<0.05$.

Among the methacrylated groups, networks MA25* displayed the lowest value of *SR* (Table 1, Figure 5). Samples MA25* UV-cured in both AcOH and HCl solutions proved to display significantly decreased *SR* with respect to the ones of respective sample MA0.3*, in line with previous gel content data. On the other hand, an opposite trend was observed in networks synthesised in PBS, whereby sample MA25(P)* displayed higher *SR* with respect to sample MA0.3(P)*, despite the comparable gel content between the two samples (Figure 4) and the significantly higher degree of functionalisation displayed by the former compared to the latter network precursor. Given that the swelling ratio was determined in PBS, the reason for this finding is likely explained by the fact that the increased molar content of methacrylamide moieties covalently-coupled to sample MA25 results in additional secondary interactions of the resulting network with PBS species, in line with the viscosity trends measured with the hydrogel-forming solutions (Table 2).

Overall, the water uptake capability of these atelocollagen networks was found to be comparable to the one of purely synthetic UV-cured PEG-based hydrogels [55] and may be exploited for the design of multifunctional wound healing devices with enhanced exudate regulation capability [8,38]. The possibility to adjust the swelling properties of this atelocollagen system by simply varying the UV-curing solution rather than by the synthesis of new photoactive precursors is appealing aiming to expand material applicability yet minimising additional time-consuming reactions.

Following characterisation of the swelling properties, the compressive properties of resulting UV-cured hydrogels were investigated. Typical stress-compression curves are reported in Figure 6, depending on the network architecture and type of UV-curing solvent employed. All samples revealed a *J*-shaped curve during compression, similarly to the case of previously-reported collagen hydrogels [30]. Networks MA25* proved to display lower values of compression at break compared to both networks 4VBC25* and MA0.3* (Figure 7), in line with the increased degree of functionalisation ($F$= 88 ± 1 mol.%) of the former network precursors, yielding highly crosslinked networks ($G$= 86±3 – 92±3 wt.%). When comparing systems with comparable degrees of functionalisation, samples 4VBC25* displayed a detectable increase in compression modulus ($E_c$= 69±8–126±46 kPa) when compared to samples MA0.3* ($E_c$= 9±1–62±13 kPa), in all environmental conditions investigated. Among the different aqueous media, UV-curing in HCl solutions was confirmed to yield networks with the lowest compression modulus (Figure 7), supporting previous rheological (Table 2) and gel content (Figure 4) measurements.

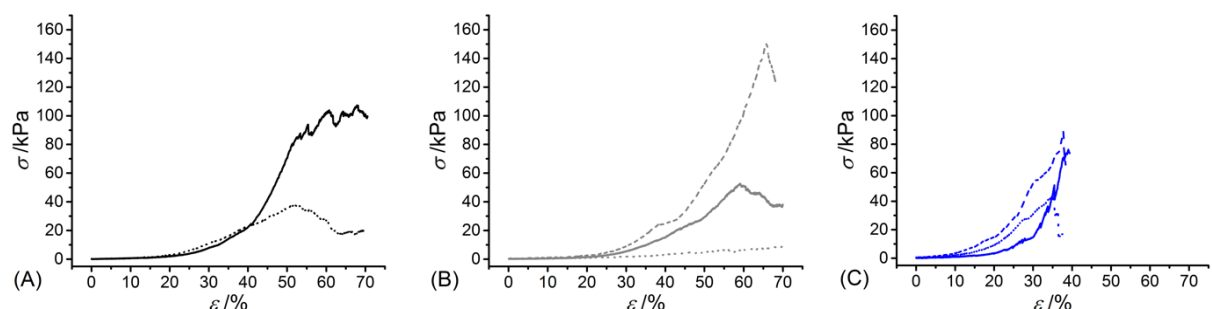

**Figure 6.** Typical stress-compression curves of hydrogels 4VBC25* (A), MA0.3* (B) and MA25* (C) prepared in either 10 mM hydrochloric acid (pH 2.1), 17.4 mM acetic acid (pH 3.4) or 10 mM PBS (pH 7.5) solution. (A): (···): 4VBC25(H)*; (—): 4VBC25(A)*. (B): (···): MA0.3(H)*; (—): MA0.3(A)*; (– – –): MA0.3(P)*. (C): (···): MA25(H)*; (—): MA25(A)*; (– – –): MA25(P)*.

Remarkably, sample 4VBC25(H)* proved to display an averaged compressive modulus of about 70 kPa, despite an averaged swelling ratio of more than 3700 wt.%. In comparison, an equivalent methacrylated network exhibited a much lower averaged compressive modulus of less than 9 kPa, supporting the low gel content ($G= 56\pm13$ wt.%) and high swelling ratio ($SR= 5202\pm401$ wt.%). The uniquely high compressive modulus *and* swelling ratio of networks 4VBC25* provides additional evidence of the key role played by the molecular stiffness of the 4VBC-based crosslinking segment, in comparison to equivalent methacrylated atelocollagen networks obtained from precursors with comparable degree of functionalisation.

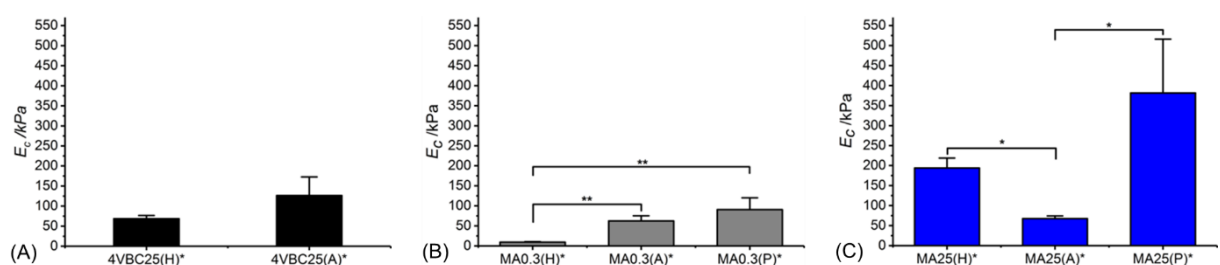

**Figure 7.** Compressive modulus ($E_c$) of atelocollagen hydrogels 4VBC (A), MA0.3 (B) and MA25 (C), prepared in varied solvents. Data are presented as mean ± SD (n=3). **: $p<0.01$, *: $p<0.05$.

When comparing systems with varied degree of functionalisation, samples MA25* proved to display significantly increased compressive modulus with respect to samples MA0.3* (and 4VBC25*), confirming the previous trends in storage modulus (Table 2). The effect of the solvent in samples MA25* proved to be less obvious, as indicated by the variation of gel content and swelling ratio. On the one hand, the significant increase in molar content of methacrylamide residues is likely to generate UV-cured networks with increased crosslink density. On the other hand, significant atelocollagen methacrylation is expected to alter the protein capability to mediate secondary interactions, potentially resulting in altered structure-function relationships. Therefore, rather than aiming at increased degree of methacrylation, these results suggest that the introduction of a limited amount of stiff 4-vinylbenzyl moieties is preferable in order to achieve defined mechanically-competent atelocollagen hydrogels.

**3.4 Enzymatic degradability**

A degradation study was carried out in enzymatic conditions in order to investigate how variations in network architecture were linked to the temporal stability of the hydrogel in near-physiologic conditions. Atelocollagen networks UV-cured in HCl solutions were selected for this investigation, due to the significant differences observed between samples 4VBC25(H)* and MA0.3(H)*. Figure 8 describes the relative mass results recorded following 4-day network incubation in an aqueous medium containing collagenase. As expected, samples of native atelocollagen revealed the highest enzymatic degradation with less than 10 wt.% remaining mass, followed by networks MA0.3H* ($\mu_{rel}$= 18±5 wt.%) and 4VBC25H* ($\mu_{rel}$= 56±5 wt.%), whilst the highest proteolytic stability was displayed by sample MA25H* ($\mu_{rel}$ = 95±1 wt.%).

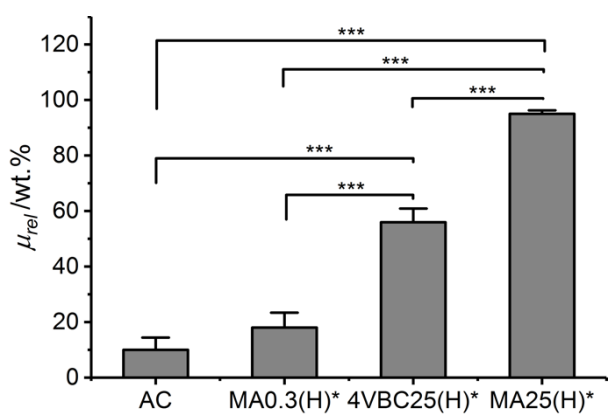

**Figure 8.** Relative mass ($\mu_{rel}$) recorded following 4-day hydrogel incubation in a collagenase-containing aqueous solution (5 CDU, 37 °C, pH 7.5). AC is the control sample of native atelocollagen. Data are expressed as mean ± SD (n=4). ***: $p < 0.001$.

Chemical crosslinking of collagen has been described as a promising strategy to increase the enzymatic stability of collagen, since the covalent crosslinks between protein molecules are insensitive to collagenases and affect the availability of tripe helical segments recognisable by enzymes [14,23,26,28,32,38]. Therefore, aforementioned degradation data successfully support the effect played by the network architecture on the enzymatic stability of the hydrogel, both in terms of type of covalently-coupled monomer as well as consequent degree of atelocollagen functionalisation. Samples MA25(H)* displayed the highest enzymatic stability given the high degree of functionalisation and gel content measured in network precursor and resulting crosslinked network, respectively. Other than that, controlled enzymatic degradation can still be achieved by simply varying the type of monomer introduced, i.e. 4-vinylbenzyl with respect to methacrylamide residue, whilst keeping the degree of functionalisation constant, in resulting photoactive precursors. The introduction of 4VBC moieties is likely to mediate secondary interactions with the exposed zinc site of active collagenases [38], so that consequent proteolytic inactivation explains the prolonged stability of sample 4VBC(H)* with respect to the equivalent variant MA0.3(H)*.

Similarly to the case of linear biopolymers [31], the manipulation of the atelocollagen network architecture is therefore proven to generate customisable hydrogels, whose degradation profiles can be combined with specific mechanical and swelling properties, depending on specific environmental (e.g. pH) and molecular parameters selected during UV-curing.

### 3.5 Cytotoxicity evaluation

Following evaluation of the enzymatic degradability, the cellular tolerability of freshly-synthesised networks 4VBC25(A)* and MA0.3(A)* was exemplarily evaluated by culturing osteosarcoma G292 in direct contact with the hydrated material. Cellular metabolism was measured by Alamar blue assay at day 1, 4 and 7 (Figure 9).

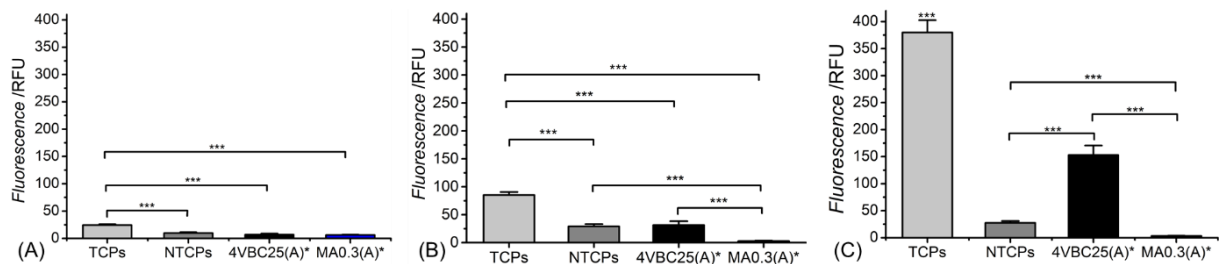

**Figure 9.** Alamar blue assay on G292 cells cultured on freshly-synthesised hydrogels 4VBC25(A)* and MA0.3(A)* at day 1 (A), 4 (B) and 7 (C). Cells seeded on non-tissue culture treated plates (NTCPs) and tissue culture treated plates (TCPs) were used as either negative or positive control, respectively. ***: $p<0.001$ (n=6).

Over the period of 7 days, G292 cells seeded on TCPs displayed significantly increased proliferation at all time points investigated. Between the group of 4VBC25(A)* and MA0.3(A)*, no significant differences was found at day 1; however, the proliferation rate of cells cultured on hydrogel 4VBC25(A)* was much higher than the one of cells cultured on hydrogel MA0.3(A)* at both day 4 and day 7. The dramatic turnover of cells on sample MA0.3(A)* was mainly due to the shrinkage of the hydrogel microstructure observed already following 1-day cell culture, in line with the fast enzymatic sample degradability measured following 4-day incubation in collagenase medium (Figure 8). Unsurprisingly, TCPs showed the highest rate of proliferation compared to 4VBC25(A)*; this observation is mainly due to the fact that the porous 3D structure of the 4VBC25(A)* hydrogel network induced low cell seeding efficiency and cellular penetration in comparison to 2D surfaces [56].

We also confirmed the cytotoxicity of the hydrogels via Live/Dead staining of G292 cells following 7-day culture (Figure 10, A and B). Almost no dead cell was found in hydrogel 4VBC(25)A*, indicating cellular tolerability in agreement with previously-discussed metabolic activity data. Besides Live/Dead staining, cool-stage SEM was also carried out on 7-day cultured hydrogel 4VBC25(A)*, as reported in Figure 10 (C and D), whereby full cellular attachment and complete surface cell coverage were observed, supporting previous findings. When the same SEM investigations were carried out on 7-day cultured hydrogel MA0.3(A)*, no cell attachment was observed (Figure S3, Supp. Inf.), in line with Alamar blue results (Figure 9) and the fast enzymatic sample degradability (Figure 8).

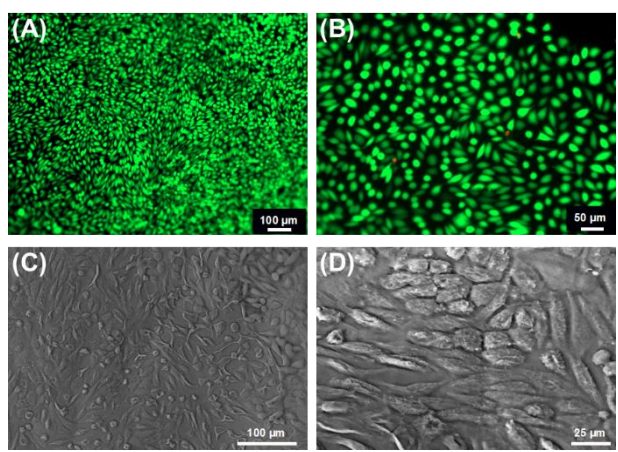

**Figure 10.** (A-B): Live/ dead staining of G292 cells following 7-day culture on hydrogels 4VBC25(A)*. (C-D): Cell attachment was confirmed by cool stage SEM on the same sample at the same time point of cell culture.

Overall, these results prove that selected functionalisation and network formation strategies did not affect atelocollagen biocompatibility, whilst proved successful in enabling systematic customisation of the pristine building block.

**Conclusion**

The introduction of either 4-vinylbenzyl or methacrylamide adducts onto type I atelocollagen proved successful to enable the synthesis of customisable UV-cured hydrogel networks, depending on the type of monomer, degree of atelocollagen functionalisation and UV-curing solution. Reaction with MA led to highly tunable degree of atelocollagen functionalisation ($F_{MA}$: 4±1-88±1 mol.%), in contrast to the reaction carried out with 4VBC ($F_{4VBC}$: 18±1 mol.%), although rapid gelation ($\tau$ = 6-12 s) was still achieved with both precursors as confirmed by photorheometry. Introduction of 4-vinylbenzyl groups proved to yield atelocollagen networks with significantly increased compression modulus ($E_c$= 69±8–126±46 kPa), storage modulus ($G'$= 73±22–390±99 Pa) and 4-day enzymatic stability ($\mu_{rel}$= 56±5 wt.%), with respect to methacrylated equivalents, due to the increased molecular stiffness of, and secondary interactions mediated by, the aromatised UV-cured crosslinking segments. Comparable variations in material properties were also observed when UV-curing atelocollagen precursors functionalised with varied content of methacrylamide functions ($F$= 19±2–88±1 mol.%), supporting the direct relationships between the degree of functionalisation of network precursors and resulting network crosslink density. The solution pH proved to affect the viscosity of respective atelocollagen solutions, whereby the monomer capability to mediate secondary interactions was found to play a role. UV-curing solutions with decreased acidity proved to generate networks with increased compression and storage modulus, as well as decreased swelling ratio, whilst no toxic cellular response was observed. These findings demonstrate the monomer-induced customisability of presented UV-cured atelocollagen hydrogel networks, whereby structure-property relationships can be controlled to enable applicability in personalised medicine and to fulfil the requirements of complex and unmet clinical needs.

**Conflict of interests**

The authors declare that there are no conflicts of interests.

**Acknowledgements**

The authors gratefully acknowledge the financial support provided by the EPSRC Centre for Innovative Manufacturing in Medical Devices (MeDe Innovation), the University of Leeds MRC Confidence in Concept scheme, the University of Leeds EPSRC Impact Acceleration Account, as well as the Clothworkers Centre for Textile Materials Innovation for Healthcare (CCTMIH). Mrs Jackie Hudson and Dr Sarah Myers are gratefully acknowledged for their kind technical assistance.

**References**

[1] Y.S. Zhang, A. Khademhosseini. Advances in engineering hydrogels. Science 2017 (356) eaaf3627

[2] D.A. Heller, Y. Levi, J.M. Pelet, J.C. Doloff, J. Wallas, G.W. Pratt, S. Jiang, G. Sahay, A. Schroeder, J.E. Schroeder, Y. Chyan, C. Zurenko, W. Querbes, M. Manzano, D.S. Kohane, R. Langer, D.G. Anderson. Modular 'Click-in-Emulsion' Bone-Targeted Nanogels. Adv. Mater. 2013 (25) 1449-1454

[3] M.D. Konieczynska, J.C. Villa-Camacho, C. Ghobril, M. Perez-Viloria, K.M. Tevis, W.A. Blessing, A. Nazarian, E.K. Rodriguez, M.W. Grinstaff. On-Demand Dissolution of a Dendritic Hydrogel-based Dressing for Second-Degree Burn Wounds through Thiol–Thioester Exchange Reaction. Angew. Chem. Int. Ed. 2016 (55) 9984-9987

[4] L. Chen, X. Yao, Z. Gu, K. Zheng, C. Zhao, W. Lei, Q. Rong, L. Lin, J. Wang, L. Jiangae, M. Liu. Covalent tethering of photo-responsive superficial layers on hydrogel surfaces for photo-controlled release. Chem. Sci. 2017 (8) 2010-2016

[5] C.M. Nimmo, M.S. Shoichet. Regenerative Biomaterials that "Click": Simple, Aqueous-Based Protocols for Hydrogel Synthesis, Surface Immobilization, and 3D Patterning. Bioconjugate Chem. 2011 (22) 2199-2209

[6] F. Koch, M. Müller, F. König, N. Meyer, J. Gattlen, U. Pieles, K. Peters, B. Kreikemeyer, S. Mathes, S. Saxer. Mechanical characteristics of beta sheet-forming peptide hydrogels are

dependent on peptide sequence, concentration and buffer composition. R. Soc. open sci. 2018 (5) 171562

[7] B.V. Sridhar, J.L. Brock, J.S. Silver, J.L. Leight, M.A. Randolph, K.S. Anseth. Development of a cellularly degradable PEG hydrogel to promote articular cartilage extracellular matrix deposition. Adv. Healthc. Mater. 2015 (4) 702-713

[8] S.E. Bulman, P. Goswami, G. Tronci, S.J. Russell, C. Carr. Investigation into the potential use of poly(vinyl alcohol)/methylglyoxal fibres as antibacterial wound dressing components. J. Biomater. Appl. 2015 (29) 1193-1200

[9] A. Aggeli, M. Bell, N. Boden, L.M. Carrick, A.E. Strong. Self‐Assembling Peptide Polyelectrolyte β-Sheet Complexes Form Nematic Hydrogels. Angew. Chemie. Int. Ed. 2003 (42) 5603-5606

[10] R. Gharaei, G. Tronci, R.P. Davies, C. Gough, R. Alazragi, P. Goswami, S.J. Russell. A structurally self-assembled peptide nano-architecture by one-step electrospinning. J. Mater. Chem. B 2016 (4) 5475-5485

[11 ] X.-H. Qin, X. Wang, M. Rottmar, B. J. Nelson, K. Maniura-Weber. Near-Infrared Light-Sensitive Polyvinyl Alcohol Hydrogel Photoresist for Spatiotemporal Control of Cell-Instructive 3D Microenvironments. Adv. Mater. 2018 (30) 1705564

[12] B.D. Fairbanks, M.P. Schwartz, A.E. Halevi, C.R. Nuttelman, C.N. Bowman, K.S. Anseth. A Versatile Synthetic Extracellular Matrix Mimic via Thiol-Norbornene Photopolymerization. Adv. Mater. 2009 (21) 5005-5010

[13] X.-H. Qin, K. Labuda, J. Chen, V. Hruschka, A. Khadem, R. Liska, H. Redl, P. Slezak. Development of Synthetic Platelet‐Activating Hydrogel Matrices to Induce Local Hemostasis. Adv. Funct. Mater. 2015 (25) 6606-6617

[14] S. Van Vlierberghe, P. Dubruel, E. Schacht. Biopolymer-Based Hydrogels As Scaffolds for Tissue Engineering Applications: A Review. Biomacromolecules 2011 (12) 1387-1408

[15] D.A. Head, G. Tronci, S.J. Russell, D.J. Wood. In Silico Modeling of the Rheological Properties of Covalently Cross-Linked Collagen Triple Helices. ACS Biomater. Sci. Eng. 2016 (2) 1224-1233

[16] M. Younesi, A. Islam, V. Kishore, J.M. Anderson, O. Akkus. Tenogenic Induction of Human MSCs by Anisotropically Aligned Collagen Biotextiles. Adv. Funct. Mater. 2014 (24) 5762-5770

[17] K. Salchert, U. Streller, T. Pompe, N. Herold, M. Grimmer, C. Werner. In Vitro Reconstitution of Fibrillar Collagen Type I Assemblies at Reactive Polymer Surfaces. Biomacromolecules 2004 (5) 1340-1350

[18] Z. Yang, Y. Hemar, L. Hilliou, E.P. Gilbert, D.J. McGillivray, M.A.K. Williams, S. Chaieb. Nonlinear Behavior of Gelatin Networks Reveals a Hierarchical Structure. Biomacromolecules 2016 (17) 590-600

[19] J.A. Burdick, G.D. Prestwich. Hyaluronic Acid Hydrogels for Biomedical Applications. Adv. Mater. 2011 (23) H41-H56

[20] G. Tronci, A.T. Neffe, B.F. Pierce, A. Lendlein. An entropy–elastic gelatin-based hydrogel system. J. Mater. Chem. 2010 (20) 8875-8884

[21] C.A. Grant, D.J. Brockwell, S.E. Radford, N.H. Thomson. Tuning the Elastic Modulus of Hydrated Collagen Fibrils. Biophys. J. 2009 (97) 2985-2992

[22] E.A.A. Neel, L. Bozec, J.C. Knowles, O. Syed, V. Mudera, R. Day, J.K. Hyun. Collagen — Emerging collagen based therapies hit the patient. Adv. Drug Deliv. Rev. 2013 (65) 429-456

[23] L.H.H. Olde Damink, P.J. Dijkstra, M.J.A. Van Luyn, P.B. Van Wachem, P. Nieuwenhuis, J. Feijen. Crosslinking of dermal sheep collagen using hexamethylene diisocyanate. J. Mater. Sci. Mater. Med. 1995 (6) 429-434

[24] A.P. Kishan, R.M. Nezarati, C.M. Radzicki, A.L. Renfro, J.L. Robinson, M.E. Whitely, E.M. Cosgriff-Hernandez. In situ crosslinking of electrospun gelatin for improved fiber morphology retention and tunable degradation. J. Mater. Chem. B 2015 (3) 7930-7938

[25] G. Tronci, R.S. Kanuparti, M.T. Arafat, J. Yin, D.J. Wood, S.J. Russell. Wet-spinnability and crosslinked fibre properties of two collagen polypeptides with varied molecular weight. Int. J. Biol. Macromol. 2015 (81) 112-120

[26] L. Duan, W. Liu, Z. Tian, C. Li, G. Li. Properties of collagen gels cross-linked by N-hydroxysuccinimide activated adipic acid deriviate. Int. J. Biol. Macromol. 2014 (69) 482-488

[27] L.H.H. Olde Damink, P.J. Dijkstra, M.J.A. Van Luyn, P.B. Van Wachem, P. Nieuwenhuis, J. Feijen. Glutaraldehyde as a crosslinking agent for collagen-based biomaterials. J. Mater. Sci. Mater. Med. 1995 (6) 460-472

[28] M.G. Haugh, C.M. Murphy, R.C. McKiernan, C. Altenbuchner, F.J. O'Brien. Crosslinking and Mechanical Properties Significantly Influence Cell Attachment, Proliferation, and Migration Within Collagen Glycosaminoglycan Scaffolds. Tissue Eng Part A. 2011 (17) 1201-1218

[ 29 ] F. Everaerts, M. Torrianni, M. Hendriks, J Feijen. Biomechanical properties of carbodiimide crosslinked collagen: Influence of the formation of ester crosslinks. J. Biomed. Mater. Res. 2008 (85A) 547-555

[30] S. Yunoki, T. Matsuda. Simultaneous Processing of Fibril Formation and Cross-Linking Improves Mechanical Properties of Collagen. Biomacromolecules 2008 (9) 879-885

[31] Y. Lee, J. Woo Bae, D.H. Oh, K.M. Park, Y.W. Chun, H.-J. Sung, K.D. Park. In situ forming gelatin-based tissue adhesives and their phenolic content-driven properties. J. Mater. Chem. B 2013 (1) 2407-2414

[ 32 ] E. Calciolari, F. Ravanetti, A. Strange, N. Mardas, L. Bozec, A. Cacchioli, N. Kostomitsopoulos, N. Donos. Degradation pattern of a porcine collagen membrane in an in vivo model of guided bone regeneration. J. Periodont. Res. 2018 (53) 430-439

[33] W.T. Brinkman, K. Nagapudi, B.S. Thomas, E.L. Chaikof. Photo-Cross-Linking of Type I Collagen Gels in the Presence of Smooth Muscle Cells: Mechanical Properties, Cell Viability, and Function. Biomacromolecules 2003 (4) 890-895

[34] I.D. Gaudet, D.I. Shreiber. Characterization of Methacrylated Type-I Collagen as a Dynamic, Photoactive Hydrogel. Biointerphases 2012 (7) 25

[35] R. Holmes, X.-B. Yang, A. Dunne, L. Florea, D. Wood, G. Tronci. Thiol-Ene Photo-Click Collagen-PEG Hydrogels: Impact of Water-Soluble Photoinitiators on Cell Viability, Gelation Kinetics and Rheological Properties. Polymers 2017 (9) 226

[36] R. Ravichandran, M.M. Islam, E.I. Alarcon, A. Samanta, S. Wang, P. Lundström, J. Hilborn, M. Griffith, J. Phopase. Functionalised type-I collagen as hydrogel building block for bio-orthogonal tissue engineering applications. J.ournal of Mater.ials Chem.istry B 2016 (4) 318-326.

[ 37 ] G. Tronci, C.A. Grant, N.H. Thomson, S.J. Russell, D.J. Wood. Influence of 4-vinylbenzylation on the rheological and swelling properties of photo-activated collagen hydrogels. MRS Advances 2016 (1) 533-538

[38] G. Tronci, J. Yin, R.A. Holmes, H. Liang, S.J. Russell, D.J. Wood. Protease-sensitive atelocollagen hydrogels promote healing in a diabetic wound model. J. Mater. Chem. B, 2016 (4) 7249-7258

[39] H. Liang, S.J. Russell, D.J. Wood, G. Tronci. A hydroxamic acid–methacrylated collagen conjugate for the modulation of inflammation-related MMP upregulation. J. Mater. Chem. B 2018 (6) 3703-3715

[40] A.K. Lynn, I.V. Yannas, W. Bonfield. Antigenicity and immunogenicity of collagen. J. Biomed. Mater. Res. B Appl. Biomater. 2004 (71) 343-354

[41] W.A. Bubnis, C.M. Ofner III. The determination of ϵ-amino groups in soluble and poorly soluble proteinaceous materials by a spectrophotometrie method using trinitrobenzenesulfonic acid. Anal. Biochem. 1992 (207) 129-133

[42] B. Starcher. A Ninhydrin-Based Assay to Quantitate the Total Protein Content of Tissue Samples. Anal. Biochem. 2001 (292) 125-129

[43] A.P. Kishan, R.M. Nezarati, C.M. Radzicki, A.L. Renfro, J.L. Robinson, M. Whitely, E. Cosgriff-Hernandez. In situ crosslinking of electrospun gelatin for improved fiber morphology retention and tunable degradation. J. Mater. Chem. B 2015 (3) 7930-7938

[44] T. Billiet, B. Van Gasse, E. Gevaert, M. Cornelissen, J.C. Martins, P. Dubruel. Quantitative contrasts in the photopolymerisation of acrylamide and methacrylamide-functionalised gelatin hydrogel building blocks. Macromol Biosci 2013 (13) 1531-1545

[45] Y. Jia, H. Wang, H. Wang, Y. Li, M. Wang, J. Zhou. Biochemical Properties of Skin Collagens Isolated from Black Carp (*Mylopharyngodon piceus*). Food Sci. Biotechnol. 2012 (21) 1585-1592

[46] G. Tronci, C.A. Grant, N.H. Thomson, S.J. Russell, D.J. Wood. Multi-scale mechanical characterization of highly swollen photo-activated collagen hydrogels. J. R. Soc. Interface 2015 (12) 20141079

[47] M. Levitt, M.F. Perutz. Aromatic Rings Act as Hydrogen Bond Acceptors. J. Mol. Biol. 1988 (201) 751-754

[48] J.D. McCall, K.S. Anseth. Thiol−Ene Photopolymerizations Provide a Facile Method To Encapsulate Proteins and Maintain Their Bioactivity. Biomacromolecules 2012 (13) 2410-2417

[49] J. Van Hoorick, P. Gruber, M. Markovic, M.R. Geert-Jan, G.M. Vagenende, M. Tromayer, J. Van Erps, H. Thienpont, J.C. Martins, S. Baudis, A. Ovsianikov, P. Dubruel, S. Van Vlierberghe. Highly Reactive Thiol‐Norbornene Photo‐Click Hydrogels: Toward Improved Processability. Macromol. Rapid Comm. 2018 (39) 1800181

[50] J. Heo, R.H. Koh, W. Shim, H.D. Kim, H.-G. Yim, N.S. Hwang. Riboflavin-induced photo-crosslinking of collagen hydrogel and its application in meniscus tissue engineering. Drug Deliv. Transl. Res. 2016 (6) 148-158

[51] L.S. Birchall, S. Roy, V. Jayawarna, M. Hughes, E. Irvine, G.T. Okorogheye, N. Saudi, E. De Santis, T. Tuttle, A.A. Edwards, R.V. Ulijn. Exploiting CH-p interactions in supramolecular hydrogels of aromatic carbohydrate amphiphiles. Chem. Sci. 2011 (2) 1349–1355

[52] F. Li, Y. Zhu, B. You, D. Zhao, Q. Ruan, Y. Zeng, C. Ding. Smart Hydrogels Co‐switched by Hydrogen Bonds and $\pi$-$\pi$ Stacking for Continuously Regulated Controlled-Release System. Adv. Funct. Mater. 2010 (20) 669-676

[53] J. Ratanavaraporn, S. Kanokpanont, Y. Tabata, S. Damrongsakkul. Effects of acid type on physical and biological properties of collagen scaffolds. J. Biomater. Sci. Polym. Ed. 2008 (19) 945-952

[54] Y. Li, A. Asadi, M.R. Monroe, E.P. Douglas. pH effects on collagen fibrillogenesis in vitro: Electrostatic interactions and phosphate binding. Mater. Sci. Eng. C 2009 (29) 1643-1649

[55] J.A. DiRamio, W.S. Kisaalita, G.F. Majetich. J.M. Shimkus. Poly(ethylene glycol) Methacrylate/Dimethacrylate Hydrogels for Controlled Release of Hydrophobic Drugs. Biotechnol. Progr. 2005 (21) 1281-1288

[56] G. Vunjak, N.B. Obradovic, I. Martin, P.M. Bursac, R. Langer, L.E. Freed. Dynamic Cell Seeding of Polymer Scaffolds for Cartilage Tissue Engineering. Biotech Progr. 1998 (14) 193-202

**Supporting information**

**Monomer-induced customisation of UV-cured atelocollagen hydrogel networks**

He Liang,[1,2] Stephen J. Russell,[1] David J. Wood,[2] Giuseppe Tronci[1,2*]

[1] Clothworkers' Centre for Textile Materials Innovation for Healthcare, School of Design, University of Leeds, United Kingdom

[2] Biomaterials and Tissue Engineering Research Group, School of Dentistry, St. James's University Hospital, University of Leeds, United Kingdom

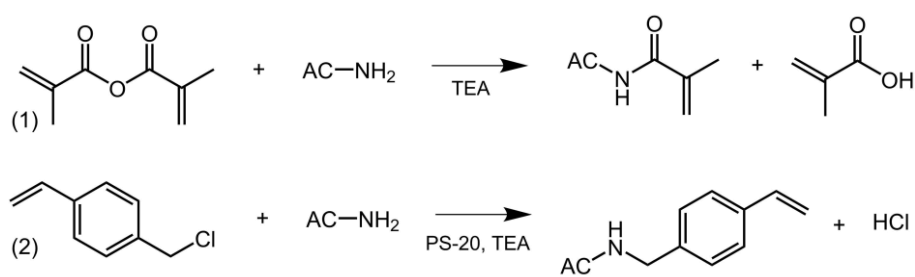

**Figure S1.** Functionalisation reaction of type I atelocollagen (AC) with either MA (1) or 4VBC (2). Both reactions proceed via amine-initiated nucleophilic substitution in the presence of trimethylamine (TEA), whereby PS-20 is used to mediate the solubility of 4VBC in water.

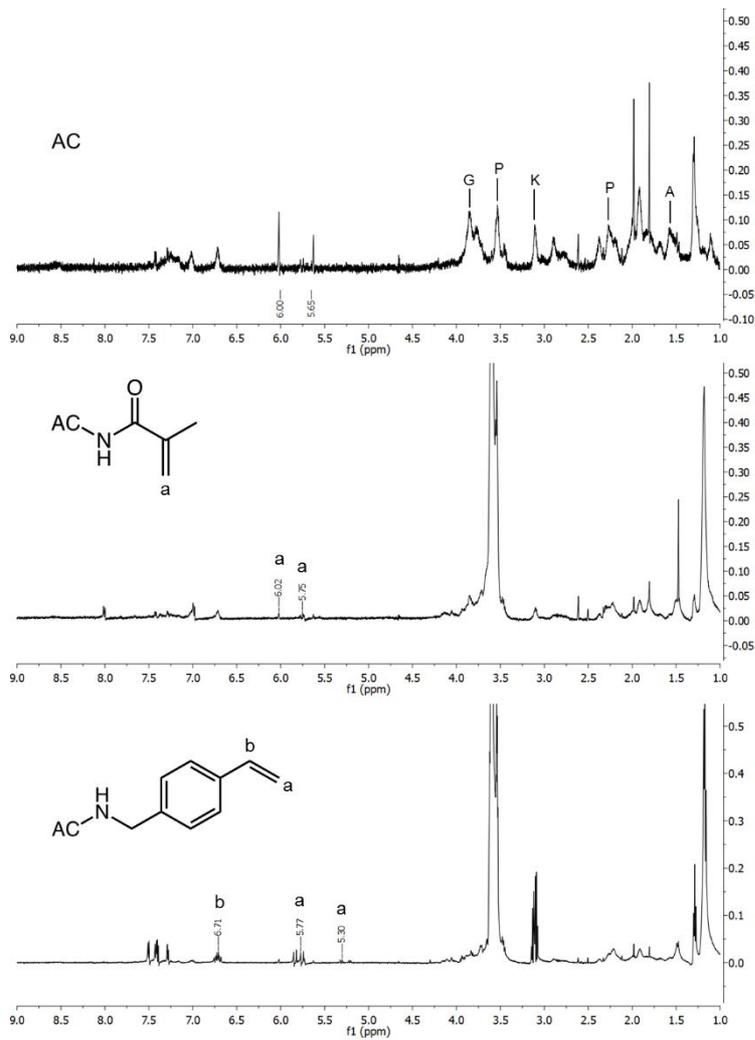

**Figure S2.** $^1$H-NMR spectra of native type I atelocollagen (AC, top), MA-functionalised AC (sample MA0.3, middle) and 4VBC-functionalised AC (sample 4VBC25, bottom), recorded in 10 mM DCl (5 mg·ml$^{-1}$) at room temperature.

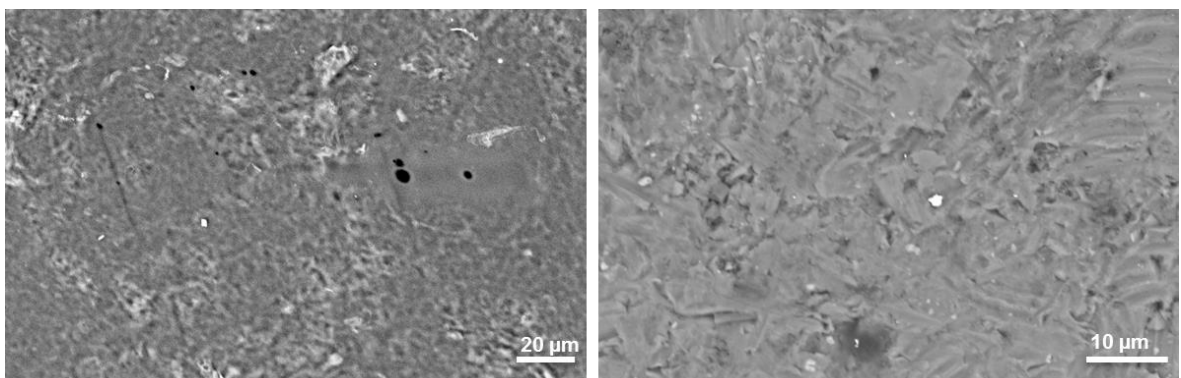

**Figure S3.** Typical cool-stage SEM image of G292 cell-seeded MA0.3(A)* hydrogel at day 7. No cells were found due to the effect of shrinkage.